
\magnification=\magstep1
\baselineskip=16pt
\vsize=8.9 true in
\hsize=6.5 true in
\hoffset=-0.1truecm\voffset=0.3truecm
\nopagenumbers\parindent=0pt
\footline={\ifnum\pageno<1 \hss\thinspace\hss
    \else\hss\folio\hss \fi}
\pageno=-1

\newdimen\windowhsize \windowhsize=13.1truecm
\newdimen\windowvsize \windowvsize=6.6truecm

\def\heading#1{
    \vskip0pt plus6\baselineskip\penalty-250\vskip0pt plus-6\baselineskip
    \vskip2\baselineskip\vskip 0pt plus 3pt minus 3pt
    \centerline{\bf#1}
    \global\count11=0\nobreak\vskip\baselineskip}
\count10=0
\def\section#1{
    \vskip0pt plus6\baselineskip\penalty-250\vskip0pt plus-6\baselineskip
    \vskip2\baselineskip plus 3pt minus 3pt
    \global\advance\count10 by 1
    \centerline{\expandafter{\number\count10}.\ \bf{#1}}
    \global\count11=0\nobreak\vskip\baselineskip}
\def\subsection#1{
    \vskip0pt plus3\baselineskip\penalty-200\vskip0pt plus-3\baselineskip
    \vskip1\baselineskip plus 3pt minus 3pt
    \global\advance\count11 by 1
    \centerline{{\it {\number\count10}.{\number\count11}\/})\ \it #1}}
\def\firstsubsection#1{
    \vskip0pt plus3\baselineskip\penalty-200\vskip0pt plus-3\baselineskip
    \vskip 0pt plus 3pt minus 3pt
    \global\advance\count11 by 1
    \centerline{{\it {\number\count10}.{\number\count11}\/})\ \it #1}}

\def\eol{\hfil\break}
\def\affl#1{\noindent\llap{$^{#1}$}}
\def\simlt{\lower.5ex\hbox{$\; \buildrel < \over \sim \;$}}
\def\simgt{\lower.5ex\hbox{$\; \buildrel > \over \sim \;$}}

{
\def\cl#1{\hbox to \windowhsize{\hfill#1\hfill}}
\hbox to\hsize{\hfill\hbox{\vbox to\windowvsize{\vfill
\bf
\cl{SHARP HI EDGES IN THE OUTSKIRTS}
\cl{OF DISK GALAXIES}
\bigskip
\cl{Edvige~Corbelli$^{1}$ and Edwin~E.~Salpeter$^{2}$}
\bigskip\rm
\cl{Preprint n.~25/93}

\vfill}}\hfill}}

\vskip5truecm
{\leftskip1.7truecm
\affl{1}Osservatorio Astrofisico di Arcetri,
\eol
Largo E.~Fermi 5, I-50125 Firenze (Italy)
\bigskip
\affl{2}Center for Radiophysics and Space Research,
\eol
Cornell University, Ithaca NY, 14853 (USA)

\vfill
To appear in The Astrophysical Journal (Dec. 10, 1993).
\vglue3truecm
}
\eject

\vglue 5 true cm
\heading{ABSTRACT}

\vglue 0.1 true in

Observations indicate that some extended outer disks have a sharp
cut off in the surface density of neutral hydrogen
when this approaches the value of $\sim 2\times 10^{19}$
cm$^{-2}$. In this paper we model these HI edges as places where
the ratio of neutral to ionized hydrogen drops rapidly due to
ionizing radiation.
We use two different models for the vertical distribution
of gas above the outer galactic plane: in the first model we
derive the density from the ideal gas law while in the second
model we insert a macroscopic pressure term and derive the
density as for an isothermal slab. We consider two different
sources of ionizing photons: external fluxes of different
intensity and spectral index due to quasars, and
a monochromatic UV flux due to neutrino decays inside and
outside the disk.
We find that galaxies which have a smaller
gas scale height, because of a higher dark matter density
or a larger external pressure, should show outer HI disks
to a lower column density and smoother HI edges.
The sharpness of the HI-HII transition and the
total column density at which the medium is 50$\%$ ionized, are
strongly correlated, irrespective of the gas or photon flux
model used.
We present several model fits to the HI sharp edge observed in
the nearby galaxy M33. If today's UV background is dominated
by attenuated quasar light, which gives $\sim 10^{-14}$ H
ionizations s$^{-1}$, a large gas scale height or
equivalently a nearly spherical halo is preferred.
If ionizing photons from decaying neutrinos are responsible
for the M33 sharp edge, then a thin outer HI disk,
and consequently a flat dark matter halo, is required.

\medskip
\noindent\underbar{\strut Subject}\ \underbar{\strut headings}:
Cosmic Background Radiation -- Galaxies:ISM --
Radio Sources: 21 cm Radiation.

\vglue\windowvsize plus 1fill minus \vsize
\eject

\pageno=1
\section{ INTRODUCTION}

The characteristics of neutral hydrogen outside the star forming
regions of disk galaxies have been outlined by Corbelli $\&$
Salpeter (1993), hereafter Paper I. In particular we have
focused on the heating inputs to the gas in low pressure conditions,
which favor non-LTE and make the HI spin temperature
deviate strongly from the kinetic temperature of the gas. Paper I
was intended for observers who have measured the HI
spin temperature through 21-cm absorption lines and want to
use it to derive information on the
cosmic background or local non-ionizing sources.
In the present paper we
study HI sharp edges discovered in spiral galaxies
(M33, NGC3198) and in surveys of high velocity clouds
(hereafter HVC). In outer disks the HI column density
decreases slowly with radius until a sharp edge
occurs at $N_{HI}=N_l\sim 2\times 10^{19}$ cm$^{-2}$ (Corbelli,
Schneider, $\&$ Salpeter 1989; van Gorkom 1991);
no residual HI is detected in emission at larger radii.
Similarly, surveys of HVC have shown that
there is a transitional column density
$N_l=5\times 10^{18}$ cm$^{-2}$ such that appreciable emission
is seen for $N_{HI} > N_l$ but
little emission is seen for $10^{18}$ cm$^{-2} < N_{HI} < N_l$
(Colgan, Salpeter, $\&$ Terzian 1990).
In Paper I we have shown that these HI edges cannot be
due to subthermal effects i.e. to a drastic depression of the
21-cm emission line when the density of the medium is lower than
a critical value and the spin temperature deviates strongly
from the gas kinetic temperature.
Here we test a totally different possibility
for HI edges i.e. that they are HI-HII transitions
due to an ionizing flux which interacts with the
neutral hydrogen distribution.

The lack of star forming regions in the outskirts of spiral galaxies
makes the cosmic background radiation extremely important
for HI ionization. This background is not directly
observable because it is easily absorbed by the thick
HI layer in the galactic plane. A careful comparison
between sensitive HI emission studies in outer regions
and theoretical results on the interaction between an ionizing
background and a gas layer, can give important information on the
spectrum and intensity of the cosmic background below 100 eV.
Several papers have
discussed this method (Felten $\&$ Bergeron
1969; Sunyaev 1969; Silk $\&$ Sunyaev 1976; Bochkarev
$\&$ Sunyaev 1977); however, due to the lack of
reliable information on low surface brightness HI and on the
mass distribution in outer regions at that time, they could not
draw any conclusion on the cosmic background.
We discuss the properties of HI edges, how they
can change from one type of galaxy to another and from one type of
object to another (from outer disks to HVC, for example).
In this paper we  focus on models which can reproduce the
characteristics of the edge which has been observed with
good sensitivity in M33.

Details on the cosmic background radiation, on the equilibrium
equations and parameters used for the gas distribution
are described in Section 2. In Section 3 we give some scaling
relation for the gas distribution around HI edges and
in Section 4 we derive numerical results for HI edges
due to several background models and give fits to the M33
sharp edge.
The characteristics of HI edges in outer disks which have local
ionizing photons from the decay of dark matter are examined in
Section 5. We summarize and discuss our results for the
cosmic UV background and
the gas distribution in the outskirts of disk galaxies in
Section 6.

\section
{EQUILIBRIUM MODELS FOR THE GAS IN OUTER REGIONS}

The ionization-recombination equilibrium depends on the photon
spectrum  and on the gas density distribution.
The UV and soft X-ray background has been reviewed in Paper I.
For the purpose of this paper however the spectrum above
0.5 keV  is not be important because we are mainly interested
in the ionizing effect of the cosmic radiation, and
photons with energy close to the Lyman limit are the most important.
As in Paper I
we use the following spectral form with two parameters
for the extragalactic spectrum between 13.6 and 500 eV:

$$ {{\hbox {d}}N \over {\hbox {d}}E} =  I\ E_{keV}^{-s}
\ {\hbox{ph}}\ {\hbox{cm}}^{-2} {\hbox{s}}^{-1} {\hbox{sr}}
^{-1} {\hbox{keV}}^{-1}\eqno(1)$$

\noindent
We shall mostly
use a steep spectrum below 500 eV, usually $s\ge1.4$.
For $s=1.4$ and $I\simeq 7.7$, eq. (1) gives the
spectral law observed above 2 keV. A background dominated by
quasar light
has a spectral index $s\simeq 2 - 3$ at zero redshift.
A constraint to the extragalactic flux comes from $H\alpha$
measurements in our Galaxy from
which one can estimate the sum of the extragalactic and local
ionizing radiation (Kutyrev $\&$ Reynolds 1989; Songalia, Bryant
$\&$ Cowie 1989). Measurements towards HVC give the
conservative upper limit of $2\times 10^5$ photons cm$^{-2}$
s$^{-1}$ for the total flux.
The spectrum suggested by Sargent {\it et al}. (1979) for the
extragalactic flux
($I\simeq 15$ and $s=2.4$) gives about $5.5\times
10^4$ photons cm$^{-2}$ s$^{-1}$ and $\zeta$ $\simeq
10^{-13}$ s$^{-1}$ for the total number of ionizations
per second for an unshielded neutral H atom (see Paper I, Section 3).
This is much lower than the value $\zeta \simeq 2.7 $
$\times 10^{-12}$ s$^{-1}$ obtained from an analysis of the
proximity effect (Bajtlik {\it et al.} 1988) at redshifts
of about 2 and fits within some evolutionary studies
of Lyman-$\alpha$ clouds
(Ikeuchi $\&$ Turner 1991). However some recent papers on flux
evolution and low redshift Lyman-$\alpha$ absorption lines estimate
lower fluxes (Madau 1992; Charlton, Salpeter, $\&$ Hogan 1993).
These models take into account the role of distant absorbers
in decreasing with time the intensity of the
cosmic radiation at energies close to the Lyman limit, and give
$\zeta$ $\sim 10^{-14}$ s$^{-1}$ at present. This corresponds
approximately to $6\times 10^{-24}$ erg cm$^{-2}$ s$^{-1}$
sr$^{-1}$ Hz $^{-1}$ at the Lyman edge,
and to eq. (1) with $I\simeq 1.4$ and $s=2.5$.
In the rest of this paper we will refer to this spectrum
as the ``Madau flux''.

The gas is considered distributed in a slab (with the
vertical extension much smaller than the horizontal one).
For a given total column density, the penetrating
flux, the gas kinetic temperature and $N_{HI}/N_{HII}$, all
change with depth inside the layer.
We therefore solve the radiative transfer, the first momentum and
energy
equations by integrating through the HI slab and iterate until
the solution converges.
Starting from the top of the slab we evaluate at each point
the attenuation and the mean energy of the flux coming from
both sides. We set the next step such that its
optical depth, at this mean energy, is 0.05 normal to the plane.
We use the ionization equilibrium and energy equations
(eq. (8) and eq. (11) of Paper I)
to find the equilibrium temperature $T_K$ at each step. Because
the mean energy of the radiation is close to the Lyman
edge we do not include HeIII and assume that
the fractional ionization of HI ($x_H$) is equal to that
of HeI when the main photon energy is above 24 eV.
The inclusion of HeIII, as well as the correct fractional ionization
of HeI, would make the HI edge slightly
sharper because high energy photons would be absorbed by HeII atoms
and re-emitted at lower energies with a higher probability of
ionizing HI atoms. However, since we will be doing a best fit
to the available HI data this would only improve the chi-square
value of the fit but it will not really make a difference in
the selection of the best-fitting model.
We use half solar abundance for metals and a helium density which
is 10$\%$ the hydrogen density.
We iterate on $x_H$ and $T_K$ at each step as well as on the
whole solution in the slab until convergence is achieved.

There are no observations so far of the vertical structure
of the gas in outer regions (at radial distances greater than
2 optical radii) or in external clouds. Even if detailed studies
of face on galaxies give evidence of broad 21-cm line profiles
in outer regions (see for example Dickey, Hanson, $\&$ Helou
1990) we don't
know yet if the linewidth is determined by microscopic thermal
motions or by bulk motion inside the disk. Therefore
we shall consider separately two sets of models for the vertical
gas distribution:

$\underline {\hbox {Model A}}$. This model assumes that
at any height $z$ above the gas midplane the gas pressure
$P(z)$ is connected to
the gas mass density $\rho(z)$ simply by the ideal gas law.
In this model we do not
include the gravity due to dark matter explicitly but
we allow the pressure above the slab, $P_{ext}=P(\infty)$,
to be a free parameter, somewhat similar to the parameter
$P_0$ in Paper I. For a given total hydrogen
(HI+HII) column density, $N_{tot}$, the vertical distribution of
the gas is then determined by the first momentum equation which
in the absence of flow reduces to the hydrostatic equilibrium
equation:

$$ {\partial P(z)\over \partial z} = 4\pi G  \rho(z)
\int_0^z \rho(z) dz \eqno (2)$$

where G is the gravitational constant.
As discussed in Paper I we expect thermal pressures
in outer regions to be much smaller than in the luminous disk
and we consider
$1 \le P_{ext}/k \le 100$ cm$^{-3}$ K ($k$ is the
Boltzmann constant).

$\underline {\hbox {Model B}}$.
In this model we omit the external pressure, but include
explicitly the gravitational force due to dark matter and allow
for a contribution to total pressure $P^{\ast}$ from
bulk motion, magnetic fields, cosmic rays, etc. We write:

$$P^{\ast}(z) \equiv \rho(z) T^{\ast} \eqno (3)$$

\noindent
where $T^{\ast}$ must exceed the gas kinetic temperature $T_K(z)$.
While $T_K(z)$ can vary with height and may change drastically
from the cold to the warm HI phase, we make the assumption
that $T^{\ast}$ can be approximated by a constant throughout the
whole slab. The ideal gas law in this case is used only to evaluate
the internal microscopic pressure. The density $\rho(z)$
is continuos and related to the
macroscopic pressure by the constant $T^{\ast}$ as
in equation (3). Here $T^{\ast}$ is in units of (cm$^2$ s$^{-2})$
but sometimes we will
express it in degrees K by multiplying it by $k/(\mu m_H)$
(with $\mu$ the mean molecular weight).
For the vertical gravity due to dark matter we use $F_z=-K z $.
The exact expression for $K$ depends on the halo eccentricity and
we present models with different values of $K$.
If we assume corotation and choose a spherical distribution
of dark matter with a power law radial density profile,
at $z<<R$ we have $K\simeq V^2/R^2$, $V$ being the asymptotic
rotational velocity.
In this case the first momentum equation along $z$ as:

$${\partial \rho(z) \over \partial z} = - \rho(z)
\Big\lbrace {K \over T^{\ast}} z +  {4\pi G
\over T^{\ast}}\int_0^z \rho(z) dz \Big\rbrace\eqno (4)$$

In Section 1 we mentioned that, beyond the first drop
associated with the optical edge,
the neutral atomic column density decreases smoothly
and slowly as one proceeds radially outward,
until it reaches a critical value $N_l$
and then drops very rapidly to unobservably small values
(Corbelli {\it et al.} 1989; van Gorkom 1991).
In Paper I we showed that this sharp outer HI edge cannot
be an artifact due to subthermal effects in emission, but there are
still two different possibilities.
The first one is that the HI drop corresponds to a drop in the
total hydrogen column density. If the gas is in stable circular
orbits, $N_{tot}=N_{HI}+N_{HII}$ can be an arbitrary function and
it could in principle have had a sharp drop since the galaxy's
formation.
The second possibility favored here and by Maloney (1992)
is that $N_{tot}$ continues to drop
only slowly below $N_l$ but the ionization-recombination equilibrium
results in $N_{HI} << N_{HII}$ for $N_{HI} < N_l$.
In the two following sections we analyze the effects of
several extragalactic background models
on the ionization structure of a gas slab and compare this with
HI observations.

\section
{THE GAS VERTICAL THICKNESS AND ITS IONIZATION STRUCTURE}

For a given extragalactic flux and total column density, $N_{tot}$,
the additional input parameter for Model A is
the external pressure, $P_{ext}$, and for Model B
the constants $K$ and $T^{\ast}$. We
assume that these quantities decrease smoothly and slowly
with increasing radial distance, so that they can be considered
constant around the region where the HI edge occurs.
We then compute the vertical structure
of the gas slab as we vary $N_{tot}$.
We define the gas vertical scale height $z_{1/2}$ as the height
above the midplane such that half of the total gas mass per unit
area lies between $+z_{1/2}$ and $-z_{1/2}$. Similarly we define
the HI scale height, $z_{HI}$.

For Model A the scale height depends not only on the gas surface
density and on the external pressure $P_{ext}$, but also
on the extragalactic flux, since this affects $T_K(z)$
and therefore the density stratification
via the ideal gas law. In the optically thin regime,
where $P_{ext}$ is the
dominant pressure term and the temperature of the medium is
uniform to a good approximation, the scale height is
$z_{1/2}=0.5 k N_{tot} T_K/P_{ext}$. For higher column
densities, however, self-gravity is no longer negligible and
the gas absorbs some of the extragalactic radiation.
The variations of the optical depth and of the gas volume density
along $z$ make the kinetic temperature vary with
height. The slab has a non-isothermal structure
with a scale
height which depends strongly on the gas total column density,
as well as on the spectrum of the incoming radiation. For this
model $z_{HI}$ inside the HI edge is very small.
If such thin slabs are warped or multilayered,
results for edges with an external ionizing flux should
not be changed appreciably, but observations may only give the
much thicker envelope instead of $z_{HI}$.
Close to the HI edge, $z_{HI}$ have larger values because
of the increased amount of warm hydrogen.

For Model B, if dark matter dominates over self gravity, the
gas density has
a gaussian vertical distribution and $z_{1/2}\simeq 0.68 \sqrt
{T^{\ast}/K}$ i.e. it
does not depend on the column density $N_{tot}$ but only on the
ratio between the ``pull down'' due to dark matter and the ``pull
up'' due to the inner macroscopic pressure.
If instead self-gravity dominates, then the gas
density is distributed as in an isothermal self-gravitating slab
(Spitzer 1942; Iba\~nez $\&$ di Sigalotti 1984) and $z_{1/2}=
0.12\ T^*/(G m_H N_{tot}) \simeq
10^{30} T^{\ast}/N_{tot}$ cm (where $m_H$ is the hydrogen mass
in grams and we have used 0.1 as helium abundance in number
with respect to hydrogen). In both cases however,
the total gas for Model B has a vertical scale height which does
not depend on the extragalactic flux but only on the local properties
of the galaxy. The HI vertical scale height depends on the
extragalactic flux but it will be a much smoother function than in
Model A.

For a spherical halo which has
$V/R\sim 6.7$ km s$^{-1}$ kpc$^{-1}$  ($K\sim 45$
km$^2$ s$^{-2}$ kpc$^{-2}$)
the ratio of dark matter to gas surface density for Model B
is about 4$\times 10^{20}/N_{tot}$.
The density distribution along $z$ near the HI edge
is close to a gaussian for most of the models we
consider since $N_{tot} \simlt 10^{20}$ cm$^{-2}$ and
dark matter dominates over self gravity. For
other models self-gravity is not completely
negligible and we always include it in our numerical analysis. The
observed variation of the thickness of the HI layer with radius
have been derived for the Milky Way by Kulkarni, Blitz, $\&$ Heiles
(1982) and most recently by Merrifield (1992). In the outermost
part ($R\simeq 20$ kpc) Merrifield's measure for the observed
vertical thickness, $h_z$,
is of order 1.3 kpc and it increases beyond as $\Delta h_z \approx
0.15 \Delta R$. If the HI distribution is assumed to ge Gaussian
then $h_z$ is the rms of the Gaussian distribution and
$z_{HI}\simeq 0.5 h_z$. An HI edge which
occurs further out, say at about 25 kpc, will have
$h_z\sim 2$ kpc and $z_{HI}\simeq 1$ kpc. A spherical halo
with a rotational velocity of about
220 km/s and $z_{1/2}\sim z_{HI}$
implies for Model B a value of $T^{\ast}\simeq 20,000 K$,
or a dispersion of about 13 km s$^{-1}$.
As might be expected this value for the parameter $T^{\ast}$ is
slightly larger than the gas kinetic temperature of warm HI or than
the HI velocity dispersion measured in face-on galaxies
(Huchtmeier $\&$ Bohnenstengel 1981; Shostak $\&$ van der Kruit 1984;
Dickey {\it et al.} 1990).
Since we don't know how the parameter $K$ (which depends on halo
eccentricity) and $T^{\ast}$ vary from galaxy to galaxy, we
shall consider models for a wide range of $K/T^{\ast}$ values.

We describe the variations of $N_{HI}$ with $N_{tot}$
using the function:

$$y\equiv {\hbox{log}}_{10} {N_{tot} \over N_{HI}}  \eqno(7)$$

\noindent
and define $N_{1/2}$ as the total column density which is
half neutral and half ionized, i.e. $y(N_{1/2})=0.3$.

Consider first a hypothetical monochromatic flux of ionizing photons
with photon energy $E$ and ionization rate $\zeta$ impinging on a
layer of total column density $N_{tot}$ (a variable) and of constant
hydrogen volume density $n$. Let $N_E$ be that neutral hydrogen
column density for which the optical depth is unity,

$$ N_E \simeq {1.3\times 10^{17}\over {\hbox {g}}_{1f}}
\Big({E_{\hbox{eV}}\over 13.6}\Big)^3
\ {\hbox {cm}}^{-2} \eqno(5)$$

\noindent
where g$_{1f}$ is the gaunt factor for photoionization from
ground level (Spitzer 1978). As in the simple
``Str\"omgren sphere theory'', we assume that the dimensionless
quantity $\zeta/(\alpha_2 n)$ is large and we define a critical
column density

$$ N_c \equiv {\zeta\over n \alpha_2} N_E \gg N_E \eqno(6)$$

\noindent
where $\alpha_2$ is the recombination coefficient excluding
captures to the first level. The neutral column density
$N_{HI}$ then varies with $N_{tot}$ as follows:
for $N_{tot} >$ (2 or 3) $N_c$, most of the hydrogen is neutral,
shielded by and outer layer with $N_{HII}\sim N_c$
and $y$ is small.
For $N_{tot}< 0.5 N_c$, on the other hand, the hydrogen is
mostly ionized with

$$N_{HI} \approx N_{tot}{\alpha_2 n \over \zeta} = {N_{tot}\over
N_c} N_E
\eqno(7)$$

\noindent
To summarize this simple model: as $N_{tot}$ drops from a little
above $N_c$ to a little below, $N_{HI}/N_{tot}$ drops rapidly.
The larger ($\zeta/n\alpha_2$) is the more rapid is the drop.
Observed values of $N_l$ where edges occur should then be
of order of $N_c$.

Realistic fluxes are non-monochromatic, and
the shape of $y$ as a function of $N_{tot}$ depends on the
shape of the photon spectrum. Large values of
$s$ in equation (1) should give results similar to that predicted
by monochromatic
photons at energies just above 13.6 eV; $N_{HI}$ predicted
outside the edge will be small and unobservable with present-day
21cm emission techniques. On the other hand for small $s$
(flat spectrum), the
variations of $y(N_{tot})$ should be more gradual, since photons of
different energies have different values of $N_E$.

For the variation of conditions with radial distance $R$ in a
realistic disk, the density $n$ varies as well as $N_{tot}$.
For Model A we assume $P_{ext}$=constant and as
$N_{tot}$ decreases radially outward, the gas pressure $P$ in
the plane decreases at first and then remains near $P_{ext}$.
The value of $y$ increases between zero and
a maximum constant value for small $N_{tot}$, $y_{max}
\approx {\hbox {log}}_{10} (\zeta/n\alpha_2)$.
For Model B as $N_{tot}$ decreases radially so does $P^{\ast}$
and $n$; $y$ continues to increase slowly for small $N_{tot}$.
Numerical results on the sharpness of HI-HII transition are
discussed in the next section.

\section
{NUMERICAL RESULTS USING THE EXTRAGALACTIC IONIZING FLUX}

\noindent
$\underline {\hbox{(a) A series of models}}$

In Table 1 we show some numerical results using
Model A (A1 to A4) and Model B (B1 to B7).
The attenuated quasar background
described by Madau is used in examining cases with
different galaxy parameters. The parameters are the external
pressure $P_{ext}$ for Model A
and the dark matter density plus the dispersion $T^{\ast}$
for Model B. As noted in the previous section,
for the range of parameters of interest in Model B
self-gravity hardly dominates over dark matter gravity.
Therefore results depend mostly on the
ratio $K/T^{\ast}$ (and not on the two parameters separately)
and we show them for three values of this ratio.
For each model we give in Table 1
the value of $N_{1/2}$, $z_{1/2}$ and $z_{HI}$;
for models where $z_{1/2}$ and $z_{HI}$ vary with $N_{tot}$
they are given for $N_{tot}=N_{1/2}$.

For the models A1, A2, A3 an increase in pressure increases
the mean volume density $n$, and consequently decreases the transition
value $N_{1/2}$ (which is of the order of $N_c$, in eq. (6)). For a
given spectrum $N_{E}$ is constant and a decrease in $N_{1/2}/N_E$
makes the HI edge less sharp.
If, for a given spectral shape, $P$ and $n$ were constant throughout
the slab then results would depend only
on the ratio $I/P$ which would take the place of
$\zeta/n\alpha_2$
in equation (6). In reality there is no exact scaling,
since the internal pressure increase in the slab depends on
$P_{ext}/N_{tot}^2$ (see eq. (2)), only when this
ratio is large the pressure is radially constant ($P=P_{ext}$).
For the Model B series at a given spectral shape we have
the approximate scaling $N_{1/2} \propto \sqrt{I z_{HI}}$ and
$z_{HI}\approx (2/3) z_{1/2} \propto \sqrt{K/T^{\ast}}$.
For Model A $z_{HI}$ is surprisingly small for the smallest
$P_{ext}$ because given the large value of $N_{1/2}$ a
small cold core forms at the center of the slab.

We also give results for different assumptions on the background:
the spectral index is the same as that given by Madau but the
intensity is ten times stronger (B4) or the background has a
spectral index different from $s=2.5$. For model B5 we use a flatter
spectrum which is an extrapolation down to 13.6 eV
of the hard X-ray background spectrum, and for model B6 we use
a steeper spectrum with $s=4$ and a photon intensity at the Lyman edge
ten times stronger than Madau's flux. In Figure 1
we show the total and HI column density as function of the
dimensionless parameter $x\equiv 20-$log$_{10}N_{tot}$ for
various models listed in Table 1.
Notice that Case A1 and B1 give very similar HI edges despite
the fact that they have different vertical scale heights.
Futhermore the
characteristics of HI edges corresponding to different spectra
but to the same $N_{1/2}$ are similar. From Figure 1 we note that
the HI column density where the curve $N_{HI}(x)$ is steepest
is very small and unobservable with present techniques; for
even smaller values of $N_{HI}$ the curve $N_{HI}(x)$ flattens out.

\vglue 0.1 true in
\noindent
$\underline {\hbox{(b) Fitting the HI edges observed in M33
and in HVC}}$

Corbelli {\it et al.} (1989) show in their fig. 10 the
second HI fall-off along the north side of the major axis of
M33. For comparison with our models, the
observed values of $N_{HI}$ should all be multiplied by
an inclination factor cos$(i)$; this factor is uncertain
and we initially assume it is close to one. Omitting
the very last point with large errors, we plot in Figures 2,
and 3 the values of $N_{HI}(R)$ at the
five outermost points (filled circles). Since the fractional change
$\Delta R/R$ across these points is small,
we assume that $P_{ext}$ and $K/T^{\ast}$ are constant across
the HI fall-off. Each model listed in Table 1 gives a unique
value of $N_{tot}$ for each
observed value of $N_{HI}$. These are shown as open circles
in figures 2 and 3. In order to compare a model with the observations
we have to make some assumptions about the spatial variation of
$N_{tot}$. We choose $N_{tot}(R)=0.1\ N_0\times 10^{R/R_l}$ and
determine $N_0$ and the scale length $R_l$ by a two parameter
least squares fit for each models. We fit log$_{10}N_{tot}(R)$
assuming an RMS error of $\pm 0.05$. This is about half the error
of the more rapidly varying log$N_{HI}$. The value of the fitted
$R_l$ and the $\chi^2$-values for the mean squared deviations
(with 3 degrees of freedom) are given in Table 1 for each model.

In the outer disk of M33 the radial fall-off of $N_{HI}$
changes drastically beyond $3\times 10^{19}$ cm$^{-2}$,
suggesting that $N_c\approx N_l\simeq 2\times10^{19}$ cm$^{-2}$
and $N_{1/2}\approx 4\times10^{19}$ cm$^{-2}$.
Table 1 and the fits in Figure 2 and 3 show indeed
that models with $N_{1/2}\simlt 3\times10^{19}$ cm$^{-2}$
give a poor fit and have large values of $\chi^2$.
They require a sudden drop
in $N_{tot}$ to fit the sharp edge i.e. if $N_{tot}$ is a smooth
function of $R$ they predict HI edges
less sharp than observed. These models are those with large
$P_{ext}$ or $K/T^{\ast}$ values (A3 and B3 in Fig. 2$(c)$
and 3$(c)$ for example) or with flat background spectra (B5 in Fig.
3$(e)$). They are rather unattractive for the face-on case
also because they
predict very small values for $R_l$ in M33. Although
observations do not give a precise value of $R_l$, we
know that $N_{HI}(R)$ before the edge occurs drops by a
factor 10 in about 20 arcmin ($\approx$ 4 kpc,
see Figure 2 of Corbelli {\it et al.} 1989). Therefore we don't
expect $R_l$ to be smaller than 20 arcmin for a face-on galaxy.
$\chi^2$ improves as $N_{1/2}$ increses.
However values of $N_{1/2}$ larger than those
shown in Table 1 require $z_{HI}>5$ kpc i.e. $K$ noticeable
smaller than estimated ($K$(M33)$\simgt40$ km$^2$ s$^{-2}$
kpc$^{-2}$) and unresonably high values of $T^*$ are needed
($T^{\ast}>10^3$ km$^2$s$^{-2}$). This is
to avoid self-gravity compression for these large column densities.

If the outer disk of M33 is not face-on but is inclined by an angle
$i$, the neutral column densities perpendicular to the galactic plane
are all smaller than those observed along the line of
sight. Assuming cos$i\simeq $0.7 or 0.5  with fluxes and
galaxy parameters as given in Table 1, the resulting total gas
scale length, $R_l$, becomes larger than for a face-on disk.
The $\chi^2$
values are slightly smaller, but as the galaxy gets closer and closer
to be edge-on, the best value of $N_{1/2}$ decreases.

With likely values of $i$ ($i\sim 45^o$) and a spectral shape like
that for quasars or steeper ($s\simgt 2$) good fits are obtained if
intensity $I$ and pressure are such that $N_{1/2}>N_l\simeq
2\times 10^{19}$ cm$^{-2}$. For the Madau flux
models with $P_{ext}\sim 15$ or $T^{\ast}/K\sim 4$
kpc$^2$ are favored since they give good fits to the sharp edge with
resonable sets of galaxy parameters for M33. For Model B a value of
$T^{\ast}\sim 100$ km$^2$s$^{-2}$ (or equivalently
$\sim 10^4$ K) is a lower limit since it implies
$K\sim 25$ km$^2$s$^{-2}$kpc$^{-2}$, too small even for a spherical
dark matter halo around M33. Slightly larger $T^*$ and $K$ values
are preferred. These models predict reasonable scale heights,
$z_{HI}\sim$ 1 or 2 kpc as observed
in the outermost parts of our Galaxy (Merrifield 1992).
If Merrifield's observations give evidence for an envelope of
an intrinsically thinner disk with warps,
vertical oscillations or multilayered slabs (clouds), then
models with $z_{HI}<1$ kpc may be more appropriate.

For HVC the shape of $N_{HI}(R)$ is unknown, but we know that
$N_l$ is smaller than
in outer disks being $N_l \sim $0.5 $\times 10^{19}$ cm$^{-2}$.
A model between A2 and A3 with $P_{ext}\sim 30$ cm$^{-3}$ K
and $z_{HI}\sim 0.3$ kpc is then a possibility.
Since HVC are located near the inner disk of our Galaxy,
larger values of $P_{ext}$ (due to coronal halo gas)
seem reasonable. The appropriate models with smaller $N_{1/2}$
predict less sharp HI edges.

\section
{HI EDGES AS PREDICTED BY DARK MATTER DECAY PHOTONS}

The dark matter decay theory recently developed by Sciama
(Sciama 1990, 1991), predicts a universe populated by
heavy neutrinos which decay radiatively producing photons
with energy $E_{\gamma}=13.6+\epsilon$
with $\epsilon \sim 1$ eV and a production rate proportional to
the density of dark matter. In this Section
we examine the possibility
that sharp HI-HII transitions in outer regions are due to these
low energy photons generated both inside and outside the gas slab.
If the local density of the gas is
sufficiently high (as in the optical region of galaxies),
these photons are produced and absorbed
locally and they represent the most important source of photons
for HI ionization. However, in considering
the effects of neutrino decay in outer regions, we have to consider
two other possible source terms.
Due to the small value of $N_{HI}$, contributions to the
local flux from low density regions surrounding the galaxy,
where photons from dark matter decay are not all absorbed
locally, might be significant.
We will refer to this as the ``halo contribution''.
Furthermore, an additional contribution is
given by the extragalactic UV photons created by the neutrino decay
in the rest of the universe (here we neglect
the effects of an additional quasar background flux).
We evaluate the source function, $\phi$ as for $R/R_{0}=3$
in the expressions given by Sciama (1990) and  we set
$\epsilon=1$ eV.
In computing the interaction of the UV flux from
radiative neutrino decay with HI at the edge of galaxies,
we solve the radiative transfer equation through the slab
taking into account
the contribution of all three components to the local
equilibrium. The external flux usually
dominates in the upper layers but its exact contribution to the
total flux in each point of the slab depends upon the external
pressure and self gravity for Model A and on the dark
matter density and velocity dispersion for Model B.

If the medium has a scale height as in Model B1 or B2
the volume density $n$ is quite small and the medium is
still fully ionized for column
densities $N_{tot}\sim 10^{20}$. This is due to
the large number of
ionizing photons or equivalently to the
inequality $n<{\sqrt{\phi/\alpha_2}}$.
Until self gravity sets in by compressing the
gas vertically there is no neutral phase. Since for
outer disks we
are interested in models in which HI is
present at the level of $5-8\times 10^{19}$ cm$^{-2}$
we ignore these highly ionized disks. We restrict
our attention to those cases with a small scale
height, i.e. a larger value of $K/T^{\ast}$, which require more
flattened dark matter halos. This means that in order
to obtain the edge at the correct column density one
needs smaller scale heights for Sciama models than for
QSO background models, or equivalently slightly larger values
of $P_{ext}$ or of $K/T^{\ast}$. In Table 1 the last five
entries show the results as we vary the galaxy parameters
(two for Model A and three for Model B). Figures 4 and 5
show $N_{HI}$ versus $x$. Models AS2 and BS3 can again be
excluded because of the small $R_l$ and large $\chi^2$.
The other three Sciama models shown in Table 1 give quite
good fits and realistic $R_l$ values. It is interesting that,
when parameters are adjusted to give the same $N_{1/2}$, the
predicted sharpness of the edge is similar for Model A and Model B
and it does not differ appreciably from what we get
using the Madau flux instead of the decaying neutrinos flux.

\section
{SUMMARY AND DISCUSSION}

In this paper and in Paper I we have discussed two types of
HI observations which
can be used to constrain the spectrum of the ionizing
radiation below 1 keV. Observations
show that neutral hydrogen in outer regions of spiral galaxies
is warm and in Paper I we have calculated the heat input required
from ionizing or non-ionizing sources. We have shown that
the ionizations from background photons, together with
collisions due to self gravity, are sufficient to
bring the spin temperature well above $T_R$ even if there are
no photons below 0.5 keV.
This means that $N_{HI}$ does not differ appreciably from
the value $N_B$, inferred from the 21-cm brightness temperature.
Therefore HI edges observed in the brightness column density
of outer disks at $N_B\simeq 2\times 10^{19}$ cm$^{-2}$
(Corbelli {\it et al.} 1989, van Gorkom 1991)
correspond to real
cut-offs in the neutral phase of the hydrogen distribution.
The observed edge column density is larger if the outer
disk is inclined.

In order to get a sharp HI edge
the intensity of the background flux needs to be stronger
than the extrapolation at low energies of the power law
observed above 2 keV. In this paper
we have modeled the extragalactic ionizing UV flux by means of
an intensity parameter $I$ and a spectral index $s$ (eq. (1)).
We have used two alternative, simplified, models to determine
the distributions of the gas pressure $P$ and the
volume density $n$ as function of height $z$ above the plane.
They have effectively only one free parameter each. In Model A
we assume that $P$ and $n$ are connected  by the ideal
gas law, involving the microscopic gas kinetic temperature
$T_K$, we do not consider dark matter gravity explicitly
and use a compression term $P_{ext}$, as free parameter.
In this model
when the gas column density is high enough, the cold
HI phase forms near the galactic plane with much larger
$n$ and smaller thickness than the warm HI and HII (for
this thin slab, dark matter gravity is indeed unimportant).
In Model B we introduce a dark matter gravity term, $K$,
and assume that the density stratification is that of an
isothermal slab which has some macroscopic temperature
$T^{\ast}$. In the range of column densities we are
interested in, results depend mostly on the single free
parameter $K/T^{\ast}$.
Model B is appropriate if outer disks are fairly homogeneous
with weak contrast between cloud structure and diffuse medium.
Magnetic pressure, cosmic ray pressure or turbulence allow
$T^{\ast}$ to be larger than $T_K$.
We assume that $P_{ext}$ (for Model A) and
$K/T^{\ast}$ (for Model B) are constant across the radial
interval where the HI edge occurs and we compute $N_{HI}$
as a function of $N_{tot}$ for different background spectra.

The ranges of spectral
indexes and flux densities of the ionizing background radiation
compatible with HI edges observations are also
compatible with $H\alpha$ measurements and evolutionary
studies of the background ionizing flux.
The spectral shape is in principle important; in practice
we find the somewhat surprising but simple result that for any
spectral index $s\simgt 1.4$  results depend rather weakly
on the spectral shape. What is more important is the
intensity of the flux at energies close to 13.6 eV which
determines the ionization rate per neutral hydrogen atom,
$\zeta$, or the parameter
$N_{1/2}$, defined as the value of $N_{tot}$ where
$N_{HI}=N_{HII}$ and roughly proportional to $\zeta/n$.
Spectra and models which give similar values of $N_{1/2}$
predict similar sharp edges. We have shown that
Model A, Model B or the decaying neutrino model
can all give reasonable fits to the HI edge observed
in M33 if their free parameter is such that
$N_{1/2}> 2\times 10^{19}$ cm$^{-2}$.
If the extragalactic flux just above
the Lyman edge is close to what Madau predicts
($s\sim 2.5$, $I\sim 1.4$), $N_{1/2}> 2\times
10^{19}$ cm$^{-2}$ implies
$P_{ext} \ll 50$ cm$^{-3}$ K for Model A, or
$z_{HI}\simgt 0.5$ kpc for Model B (as for an almost spherical
dark matter halo). For a galaxy with a modest inclination
respect to our line of sight as $N_{1/2}$ gets larger the fits
improve. Models which give $N_{1/2}\sim$ 3 or 4 $10^{19}$ cm$^2$
fit the M33 edge considerably well and require a reasonable
set of galaxy parameters.

As $P_{ext}$ or $K/T^{\ast}$ decreases, the transition column
density, $N_{1/2}$,
and the sharpness of the edge increase. In the case of a
quasar background, the number of ionizing photons
coming from distant
extragalactic sources is likely to be constant.
Galaxies with a higher pressure, i.e
with a substantial amount of coronal gas or dark matter in
outer regions, should then present a smaller value of $N_{1/2}$ and
smoother edges. Similarly,  HI column densities
smaller than $N_{1/2}$ can survive in HVC if they have a vertical
scale height smaller than that of outer disks due to
stronger compression (from  $P_{ext}$ or local dark matter).
With the
observational sensitivity available at the moment, this
may have led to a selection bias: edges which occur at larger
HI column densities are
easier to observe and the observed sharpness may then be larger
than the average. The galaxy
NGC 3198 shows an HI edge rather similar to that of M33.
This is not too surprising since they might have similar
dark matter densities (for spherical halos $V/R\sim 5-6$ km/s
kpc$^{-1}$ for both galaxies). However if the dark matter
density changes from one type of galaxy to another, then
we expect corresponding changes in the HI edge.

If decaying neutrinos provide the UV ionizing photons then
the HI scale height must be quite small ($z_{HI}< 0.5$
kpc) for the gas to stay
mostly neutral at $N_{tot}\sim 10^{20}$ cm$^{-2}$. However
once this small scale height is provided, by using for
example a flat dark matter halo or a low $T^{\ast}$,
the sharpness of the HI
edge is rather similar to that predicted by a quasar
background flux which gives the same $N_{1/2}$ value.
For the outer disk of our Galaxy $z_{HI} > 0.5$ kpc,
however there are no observations of the HI vertical scale
height just inside
a sharp HI edge. These observations would be of great
interest
for selecting the most appropriate model
(decaying neutrinos, Model A or Model B) for outer disks with
a sharp HI edge.
We hope that in the near future detailed observations of other
HI edges will become available together with measurements of
the HI spin temperature and its vertical scale height
just before the edge occurs.

\heading{ACKNOWLEDGMENTS} We are grateful to J. Charlton, P. Lenzuni,
R. Reynolds, M. Roberts, D. Sciama, J. H. van Gorkom and to the
referee for useful comments to the original manuscript.
This work was supported in part by NSF grant AST 91-19475,
INT 89-13558, and by the Agenzia Spaziale Italiana.

\vfill
\eject

\vglue 0.1 in

Table 1

\vfill
\eject

\def\refindent{\advance\leftskip by 24pt \parindent=-24pt}

\heading {REFERENCES}

\refindent
Bajtlik, S., Duncan, R. C., $\&$ Ostriker, J. P. 1988, ApJ, 327, 570.\par

Bochkarev N. G. $\&$ Sunyaev R. A. 1977, Sov. Astr., 21, 542.\par

Charlton, J. C., Salpeter, E. E., $\&$ Hogan, C. J. 1993, ApJ, 402, 493.\par

Colgan, S. W. J., Salpeter, E. E., $\&$ Terzian, Y. 1990, ApJ, 351, 503.\par

Corbelli, E., $\&$ Salpeter, E. E. 1988, ApJ, 326, 551.\par

Corbelli, E., $\&$ Salpeter, E. E. 1992 (Paper I), ApJ, submitted.\par

Corbelli, E., Schneider, S. E., $\&$ Salpeter, E. E. 1989, AJ, 97, 390.\par

Dickey, J. M., Hanson, M. M., $\&$ Helou, G. 1990, ApJ, 352, 522.\par

Felten, J. E. $\&$ Bergeron, J. 1969, Astrophys. Lett., 4, 155.\par

Huchtmeier, W. K. $\&$ Bohnenstengel, H. D. 1981, A$\&$A 100, 72.\par

Iba\~nez, S. M. H., $\&$ di Sigalotti, L. 1984, ApJ, 285, 784.\par

Ikeuchi, S., $\&$ Turner, E. L. 1991, ApJ 381, L1.\par

Kulkarni, S. R., Blitz, L., $\&$ Heiles, C. 1982, ApJ, 259, L63.\par

Kulkarni, S. R., $\&$ Heiles, C. 1988, in Galactic and Extragalactic
Radio Astronomy,
ed. G. L. Verschuur and K. I. Kellerman (New York:Springer-Verlag),
p. 95.\par

Kutyrev, A. S., $\&$ Reynolds, R. J. 1989, ApJ, 344, L9.\par

Maloney, P. 1992, ApJ, submitted.\par

Merrifield, M. R., 1992, AJ, 103, 1552.\par

Sargent, W. L. W., Young, P. J., Boksenberg, A., Carswell, R. F.,
$\&$ Whelan,
J. A. J. 1979, ApJ, 230, 49.\par

Sciama, D. W. 1990, ApJ, 364, 549.\par

Sciama, D. W. 1991, A$\&$A, 245, 243.\par

Shostak, G. S., $\&$ van der Kruit, P. C. 1984, A$\&$A, 132, 20.\par

Silk, J., $\&$ Sunyaev, R. A. 1976, Nature, 260, 508.\par

Songalia, A., Bryant, W., $\&$ Cowie, L. L. 1989, ApJ, 345, L71.\par

Spitzer, L. 1942, ApJ, 95, 329.\par

Spitzer, L. 1978, Physical Processes in The Interstellar
Medium, (Wiley: New York).\par

Sunyaev, R. A. 1969, Astrophys. Lett., 3, 33.\par

van Gorkom, J. H. 1991, ASP Conference series n.16, (proceedings
3rd Haystack Observ. Conference on Atoms, Ions and Molecules,
ed. A. D. Haschick and P. T. P. Ho), p. 1.\par

\vfill
\eject

\heading
{FIGURE CAPTIONS}

{\bf Figure 1}. $N_{tot}$ (solid line) and $N_{HI}$
(dashed lines) as functions of the dimensionless parameter
$x\equiv 20-{\hbox {log}}_{10}N_{tot}$ for models as given in
Table 1. The filled triangles indicate the HI column density
at $N_{tot}=N_{1/2}$. In 1$(a)$ we show cases for Model A and in
1$(b)$
for Model B. The small dashed line refer to model A1 in 1$(a)$
and B1 in 1$(b)$,
the medium dashed to model A2 and B2, and the large dashed to model
A3 and B3.
The dot-dashed line refer to model A4 and B4 in 1$(a)$ and
1$(b)$ respectively, and have a different
background intensity than used in A1,A2,A3.
In 1$(c)$ the spectral slope differs from 2.5 and the models shown
are B5 (large dashed line), B6 (medium dashed line) and B7
(dot-dashed line). B6 and B7 have the same $N_{1/2}$.

{\bf Figure 2}. The
values of $N_{HI}$ observed near the edge in M33
(filled circles) and the
corresponding total gas density $N_{tot}$ (open circles)
for a given Model A. Models shown are labeled as in Table 1.
The straight line is the best linear fit
to log$N_{tot}(R)$.

{\bf Figure 3}. The
values of $N_{HI}$ observed near the edge in M33
(filled circles) and the
corresponding total gas density $N_{tot}$ (open circles)
for a given Model B. Six models are shown and labeled as
in Table 1. The straight line is the best linear fit
to log$N_{tot}(R)$.

{\bf Figure 4}. As in Figure 1 we plot $N_{tot}$ (solid line)
and $N_{HI}$ (dashed lines) as functions of the dimensionless $x$.
Here we used a flux as given by the neutrino decay
at 3 solar radii (internal and external). The filled triangles
indicate the HI column density
at $N_{tot}=N_{1/2}$. In $(a)$ we plot the HI
column density for Case AS1 (small dashed line) and AS2 (medium
dashed line) as given in Table 1. In $(b)$ we plot the HI
column density for Case BS1 (small dashed line), BS2 (medium
dashed line) and BS3 (large dashed line).

{\bf Figure 5}. The
values of $N_{HI}$ observed near the edge in M33
(filled circles) and the
corresponding total gas density $N_{tot}$ (open circles) for
four decaying neutrinos models described in Table 1.

\end